\documentstyle[12pt,a4wide,epsf]{article}
\begin{document}

\begin{titlepage}

\begin{flushright}
HD--THEP--96--47\\
hep-ph/9611250\\
\end{flushright}
\quad\\
 \vspace{1.8cm}
\begin{center}
{\bf\LARGE Gluon-mass effects in quarkonia decays,\\ 
$e^+e^-$ annihilation and the scalar glueball current}\\
\vspace{1cm}
Jue-ping Liu\\
{\normalsize\it Department of Physics, Wuhan University}\\
{\normalsize\it Wuhan/Hubei 430072, P.R. of China}\\
\bigskip
Werner Wetzel\\
{\normalsize\it 
Institut f\"ur Theoretische Physik, Universit\"at Heidelberg}\\
{\normalsize\it Philosophenweg 16, D-69120 Heidelberg,Germany}\\
\bigskip
November 4, 1996\\
\end{center}
\vspace{2.0cm}

\begin{abstract}\noindent
In this paper we continue previous efforts in the literature
to determine phenomenological
values for the gluon mass by confronting theoretical results
obtained in a theory of massive gluons with experimental values
or results directly referring to the nontrivial structure
of the Yang-Mills vacuum, e.g. to the presence of the gluon condensate. 
The 
decays of heavy quarkonia into $3$ gluons and $2$ gluons + photon
are considered in detail as well as the correlators of the 
electromagnetic current and the scalar glueball current.
Based on the analysis for the latter quantities a value for the gluon mass
in the range of 500-600 MeV is estimated from the 
standard SVZ-value of the gluon condensate.
\end{abstract}

\end{titlepage}
\newpage

\section{ Introduction}

Starting with the original suggestion by Cornwall \cite{Cor1}, the 
possibility of a nonzero gluon mass has been discussed in various 
contexts over the past 15 years. It has been argued (c.f. \cite{Cor2}) 
that the severe infrared singularities of the gluon interaction might 
resolve themselves in the creation of a gluon mass. The emergence of 
this mass might be closely related to the nontrivial nature of the QCD 
vacuum, in particular to the gluon condensate of Shifman, Vainshtein and Zakharov \cite{SVZ}.

However unlike the by now very much established gluon condensate or 
the constituent mass of quarks, the concept of a gluon mass meets a 
lot of reservation as it touches the fundamental principles of gauge 
invariance and renormalizability.

As argued by many authors and in particular by \cite{LLS}, the only way to 
make massive nonabelian vector mesons compatible with a decent high 
energy behaviour, i.e. with renormalizability, seems to be the Higgs 
mechanism. The effective Lagrangian advocated in \cite{Cor2} for a massive gluon
theory \footnote{note, that we do not follow 
the suggestion of Cornwall to use the classical equation of motion to 
express the nonlinear $\sigma$-field in the Lagrangian in terms of the gauge 
field} in fact 
can be understood as a Higgs model with the ramification that all 
physical Higgs particles have been removed from the theory by making 
them infinitely heavy. More precisely, this model is the 
generalization of the $SU(2)$ Higgs-model with a frozen 
radial degree of freedom to the case of the gauge group $SU(3)$. In this 
gauged nonlinear $\sigma$-model, the $8$ would-be Goldstone bosons 
provide the longitudinal polarization degrees of freedom for the 
gluons. In the unitary gauge, they are absorbed into the redefinition 
of the gauge field, leaving us with a massive vector boson theory with 
a global $SU(3)$ symmetry.

Though the conflict between a gluon mass and gauge invariance is 
avoided in this model, the renormalizability has obviously been lost 
in the way of going from a linear $\sigma$-model to the nonlinear one. 

One of the main questions then is, how severe the loss of 
renormalizability really is. From the situation of $SU(2)$, where the 
resulting new divergence is only logarithmic in the cut-off or the 
physical Higgs mass - a fact known as screening theorem of Veltman 
\cite{Velt} -, we may hope for a similar behaviour in the $SU(3)$ case. To 
clarify this point, the systematic analysis of  \cite{Appl} for the 
divergencies should be extended from $SU(2)$ to $SU(3)$.

Another critical issue to the significance of this model as an 
effective low energy theory is the confinement property. In the Higgs 
phase, i.e. the phase corresponding to superconductivity, magnetic 
monopoles are confined and the Wilson loop in the fundamental 
representation satisfies a perimeter law. It has been argued in \cite{Ak} 
that at some critical coupling the model is expected to undergo a 
transition to a phase with condensed vortices. In this phase the 
't Hooft loop, i.e. the order parameter related to  confinement of 
monopoles, can be shown to satisfy a perimeter law implying - 
according to the classification of phases by 't Hooft \cite{tH1} - an area 
law for the Wilson loop in the fundamental representation.

On the phenomenological side, effects of a finite gluon mass have been 
considered in the past for the standard quarkonium decays into gluons 
\cite{Par}, potential models for glue-balls \cite{Soni}, 
approximations to the QCD-Hamiltonian combined with a variational method
\cite{Szcz},  
a possible relation between 
the gluon condensate and the gluon mass through QCD sum-rules \cite{Graz}, 
the two-gluon exchange contribution to the Pomeron \cite{Halz}, and the 
vacuum functional of $SU(3)$ Yang-Mills theory \cite{Kog}.

In a recent analysis \cite{Cons} of $J/\psi$ and $\Upsilon$ decays it was 
investigated in detail whether the deficiencies between 
experiment and theory for the total decay rate and the inclusive 
photon spectrum can be understood as being due to relativistic 
corrections or alternatively due to a finite gluon mass. The conclusion of 
\cite{Cons} is 
that the relativistic corrections needed to bring experiment and theory into agreement
do not scale in the expected manner when going from c-quarks to b-quarks.
The alternative with a gluon mass leads to values of $.66 GeV$ and $1.17 GeV$ for 
$J/\psi$ and $\Upsilon$ decays, respectively. The authors interpret the rise of the gluon mass parameter as an indication that without phase-space limitation 
the true gluon mass would be of the order of $1.5 GeV$ and is reduced to effectively
smaller values due to phase space limitations in $\Upsilon$ and $J/\psi$ decays.

As a continuation of previous efforts to determine phenomenological
values for the gluon mass, we treat in this paper the following 
three subjects. First we complete in section 2 the analysis of \cite{Cons} by
taking into account a finite gluon mass not only in the phase
space but also in the amplitudes for $J/\psi$ and $\Upsilon$ decays.
For this purpose we derive the analytical form of the Dalitz distributions 
for $(Q\bar{Q})_1^3 \rightarrow 3g$,  
$(Q\bar{Q})_1^3 \rightarrow \gamma + 2g$ and $(Q\bar{Q})_0^1 \rightarrow 2g$
for massive gluons to leading order in 
$\alpha_{s }$, thus extending the classical formulas originally derived for positronium decays (e.g. ref. \cite{LandLif}). Using these results we 
determine by a comparison with the results of
\cite{Cons} corrected values for the gluon mass.

As a further basic process affected by a finite gluon mass we consider 
in section 3 the $e^+e^-$ annihilation into mass-less quarks. The analytic form of the vacuum polarization to order $\alpha_s$ with a massive vector boson has been known for a long time \cite{Wet}. Analyzing the 
dispersion relation
for the derivative of the vacuum polarisation function, a 
systematic expansion in powers of $M^2/Q^2$ can be derived for 
the Euclidean region. This expansion can serve as a basis to relate the 
gluon mass M to the gluon condensate by equating the $1/Q^4$ term in 
this expansion with the $1/Q^4$ term in the operator product expansion as used before in ref. \cite{Graz}. As it turns out, the expression obtained in this way
is different from the expression in ref. \cite{Graz}, where a
Mellin-transform technique has been used to isolate the $1/Q^4$
term in the loop diagrams. Our relation can be used to derive a simple
expression for the gluon mass in terms of the gluon condensate.

In a similar spirit we deal in section 4 with the two-point function for
the scalar glue-ball current instead of the electromagnetic current.
Matching the result obtained for this two-point function in the massive
theory with the result obtained from the operator expansion in the
massless theory, a similar, albeit not exactly identical, relation
between the gluon mass and the gluon condensate is obtained.

In section 5 finally the results are summarized and a few 
remarks concerning possible future directions of this research are made.

\section{ Heavy quarkonia decays into gluons}

In this section we present our results for the decay rates of
triplet $(Q\bar{Q})_1^3$ and singlet $(Q\bar{Q})_0^1$
$S$-wave quarkonia states into massive 
gluons. The calculation uses the standard approximation of neglecting 
the dependence on the momenta of the heavy quarks. Accordingly we can write:

\begin{equation}
d\Gamma( (Q\bar{Q}) \rightarrow ..)  \sim |\psi(0)|^2 d\Gamma( Q 
+ \bar{Q} \rightarrow ..)
\end{equation}
\noindent
where $\psi(0)$ represents the spatial $Q\bar{Q}$ wave-function at the origin and 
$d\Gamma( Q + \bar{Q} \rightarrow ..)$ is calculated with $Q$ and $\bar{Q}$ at rest.

We use $\eta = (M/2m)^2$ to parametrize the dependence on the ratio
between gluon mass $M$ and quark mass $m$. On the level of our approximation, $2m$ of course can be 
identified with the mass of the respective singlet or triplet 
quarkonium state.

We start with the $S_0^1 \rightarrow 2g$ decay rate which is given by:

\begin{equation}
\label{gg}
\Gamma = \Gamma_0 ((1-4\eta)/(1-2\eta)^2 )(1-4\eta)^{1/2} 
\end{equation}

where $\Gamma_0$ is the decay rate for $M=0$ and the gluon mass effects on 
the amplitude squared and the phase space have been separated for convenience. 
As can be seen from eq.(\ref{gg}) the amplitude 
squared decreases with increasing $M$, though compared to the phase space the effect only starts at one order higher in $\eta$ .

For the Dalitz distribution of  $^3S_1 \rightarrow ggg$  we obtain the basic formula:

\begin{eqnarray}
\label{ggg}
\lefteqn{d\,\Gamma(^3S_1 \rightarrow ggg)  = } \\ \nonumber
& & \,{{\Gamma _{0}}\over {\left( -9 + {{\pi }^2} \right)}}\,{{\delta (-2\,\left( 1 - 3\,\eta  \right)  + x_{1} +
  x_{2} + x_{3})\,dx_1\,dx_2\,dx_3}\over {{{x_{1}}^2}\,{{x_{2}}^2}\,{{x_{3}}^2}}}\,\lbrack \\ \nonumber
& &\hspace{0pt}\eta\,{16\over 3}\,{{\left( 1 - 3\,\eta  \right) }^2} \,\left( 1 -
  {{51\,\eta }\over 8} - {{15\,{{\eta }^2}}\over 4} \right) \\
  \nonumber
& &\hspace{0pt}+\left( {{x_{1}}^2} + {{x_{2}}^2} + {{x_{3}}^2} \right)\,\left( 1 - 14\,\eta  + 48\,{{\eta }^2} + 25\,{{\eta }^3}
  \right)   \\
  \nonumber
& &\hspace{0pt}-2\,\left( {{x_{1}}^3} + {{x_{2}}^3} + {{x_{3}}^3} \right)\,\left( 1 - {{17\,\eta }\over 3} - 3\,{{\eta }^2} \right)  \\ \nonumber
& &\hspace{0pt}+\left( {{x_{1}}^4} +
  {{x_{2}}^4} + {{x_{3}}^4} \right)\,\left( 1 + {{\eta }\over 2} \right) \,\rbrack \\ \nonumber
\end{eqnarray}
\noindent
where the variables $x_i$  are related to the gluon energies $E_i$ by
$x_i= E_i /m - 2\eta$.
Similar to before $\Gamma_0$ represents the integrated decay rate for mass-less gluons.

Apart from the constant term proportional to $\eta$  , eq.(\ref{ggg}) has a 
similar form as in the mass-less case, i.e. the numerator of the 
distribution can be written as $(x_1^2 P(x_1) + cycl)$, 
where $P(x)$ is a second order polynomial in $x$. For $\eta=0$, $P$ reduces
to the form $P(x)=(1-x)^2$, well known from the classic positronium formula.

When one of the gluons, say gluon 1, is replaced by a photon, the corresponding
expression reads with $z=E_{\gamma}/m$ replacing $x_1$:
  
\samepage{
\begin{eqnarray}
\label{gamma2g}
\lefteqn{d\,\Gamma(^3S_1 \rightarrow \gamma gg)  = } \\ \nonumber
& & \,{{\Gamma _{0}}\over {\left( -9 + {{\pi }^2} \right)}}\,{{\delta (-2\,\left( 1 - 2\,\eta  \right)  + z +
  x_{2} + x_{3})\,dz_1\,dx_2\,dx_3}\over {{{z}^2}\,{{x_{2}}^2}\,{{x_{3}}^2}}}\,\lbrack \\ \nonumber
& &\hspace{0pt}+{8\over 3}\,\eta\,\left( 1 - {{25\,\eta }\over 2} \right) \,{{\left( 1 - 2\,\eta  \right) }^2}\, \\ \nonumber
& &\hspace{0pt}+z\,32\,\left( 1 - 2\,\eta  \right) \,\left( 1 - {{\eta }\over 4}\right) \,{{\eta }^2} \\ \nonumber
& &\hspace{0pt}+ {{z}^2}\,\left( 1 - 2\,\eta  \right) \,\left( 1 - 6\,\eta  - 6\,{{\eta}^2} \right) \\ \nonumber
& &\hspace{0pt}-2\,{{z}^3}\,\left( 1 - {{10\,\eta }\over 3} - 2\,{{\eta }^2} \right) \\ \nonumber
& &\hspace{0pt}+{{z}^4}\,\left( 1 + {{\eta }\over 2} \right)  \\
  \nonumber
& &\hspace{0pt}+\left( {{x_{2}}^2} + {{x_{3}}^2} \right)\,\left( 1 - 8\,\eta  + 22\,{{\eta }^2} + 8\,{{\eta }^3} \right)  \\ \nonumber
& &\hspace{0pt}-2\,\left( {{x_{2}}^3} + {{x_{3}}^3} \right)\,\left( 1 - {{10\,\eta }\over 3} - 2\,{{\eta }^2} \right)
    \\ \nonumber
& &\hspace{0pt}+\left( {{x_{2}}^4} + {{x_{3}}^4} \right)\,
\left( 1 + {{\eta }\over 2} \right)   \\ \nonumber
& &\hspace{0pt}-\left({{x_{2}}^2}\,{{x_{3}}^2} \right)\, \eta \,\rbrack \\ \nonumber
\end{eqnarray}
}
 
Using these expressions and integrating over phase space we can now make a simple analysis to estimate new values for the gluon mass which take into account both phase-space and amplitude effects. The results are shown in Figs. 1,2, where the integrated decay rate is plotted as a function of $M^2/4m^2$.
As can be seen, the inclusion of the gluon mass in the amplitude leads to
an increase of the decay rate. This feature might be expected on the basis that
for massive gluons there are in addition the channels involving longitudinal
gluons.
Using the values for the gluon mass from ref.(\cite{Cons}) i.e. $M=.66 GeV$ and
$M=1.17 GeV$ for $J/\psi$ and $\Upsilon$, respectively, we conclude that in order to reach the same level of suppression the gluon mass values change to
$M=.76(.79) GeV$ and $M=1.41(1.4) GeV$, where the numbers in brackets refer to the case where the $\gamma+2g$ reaction is used to make the matching.

\begin{figure}[p]
\label{F3g}
\epsfxsize=12cm
\epsffile{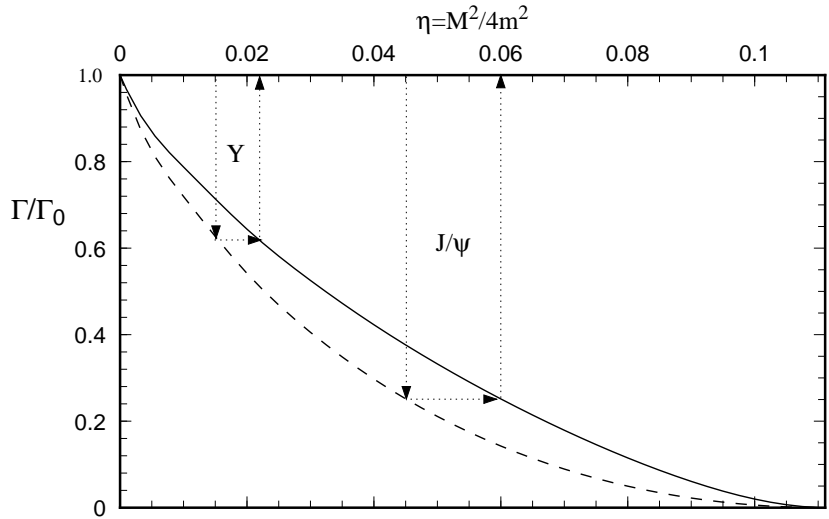}
\caption{$^3S_1 \rightarrow ggg$ decay rate as a function of $\eta=M^2/4m^2$, normalized to the decay rate for mass-less gluons. The dashed line represents
the case when the gluon mass is only taken into account in the phase space.
The dotted lines indicate how the gluon mass has to be changed to reach the same suppression when taking the gluon mass into account also in the amplitude}
\end{figure}

\begin{figure}[p]
\label{Fgamma+2g}
\epsfxsize=12cm
\epsffile{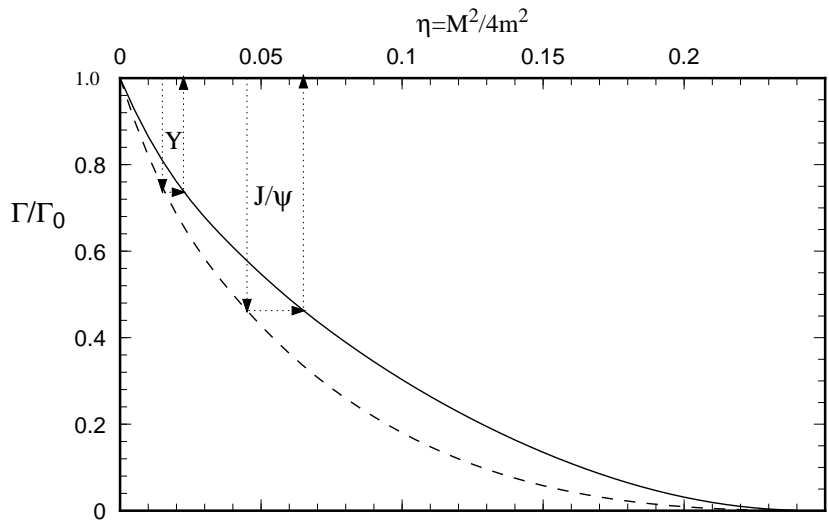}
\caption{The analogue of Fig. 1 for $^3S_1 \rightarrow \gamma+gg$}
\end{figure}

\bigskip
Another quantity of interest is the photon energy spectrum.
The analytic form is given in the Appendix. Examples for the effect
of the gluon mass on the shape of this distribution are shown in Fig. 3. As can be seen, the larger gluon mass needed to produce
the same amount of suppression when the gluon mass effect is taken
into account in both the amplitude and the phase space, leads to a sharper
cut-off at the end of the spectrum. However, in this region we cannot take
the massive gluon model really serious, and there are other
effects like the Fermi-motion of the quarks to which this behaviour
is sensitive to.

\begin{figure}[p]
\label{Fgammadist}
\epsfxsize=12cm
\epsffile{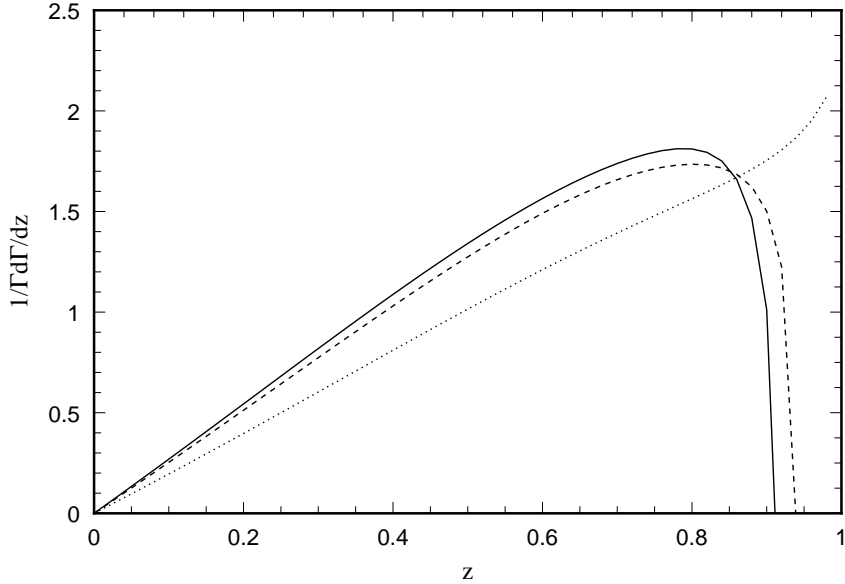}
\caption{The normalized photon energy spectrum for $\Upsilon \rightarrow \gamma+gg$ as a function of $z=E_{\gamma}/m$. The pair of gluon mass
values chosen, i.e. $M=1.41GeV$ and $M=1.17GeV$, are the ones determined
above to reproduce the experimental suppression of the total decay rate. As before, the dashed line is the curve involving the phase-space effect only.
For comparison also the spectrum for zero gluon mass is shown (dotted line)}
\end{figure}

\section{ $e^+e^-$ annihilation}

\begin{figure}[p]
\label{Fvacpol}
\epsffile{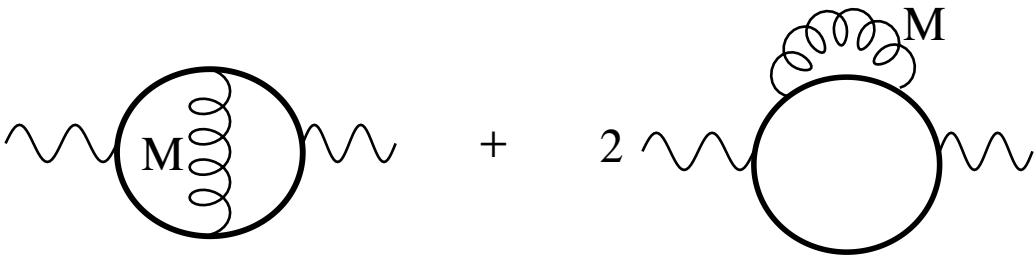}
\caption{The leading order diagrams with a gluon of mass $M$}
\end{figure}

As a further basic process we consider now the leading order effect
of a massive gluon on the electro-magnetic vacuum polarisation, i.e. the
diagrams given in Fig. 4. To simplify matters, we take the massless quark limit. A systematic expansion in terms of $M^2/Q^2$ can most easily be derived starting from the dispersion relation
for the derivative of the vacuum polarisation function.

\begin{equation}
\label{disprel}
{\rm D}({Q^2}) = {Q^2}\,\left( \int _{0}^{\infty }ds\,{{R(s)}\over {{{\left(
  {Q^2} + s \right) }^2}}} \right)
\end{equation}
\noindent
The imaginary part $R$ for the diagrams in Fig. 4
has been known for a long time \cite{Wet}, and for $ x=M^2/s < 1$ is given by:
 
\begin{eqnarray}
\label{vacpol:1}
\lefteqn{R(x) = } \\ \nonumber
& &{{N_c^2-1} \over {4\,N_c}}\,(
{3\over 2} - 2\,x - 5\,{x^2} + 2\,x\,\ln (x) + 3\,{x^2}\,\ln (x) \\
  \nonumber
& &\hspace{2.6 cm}-2\,{{\left( 1 + x \right) }^2}\,\left( \ln (x)\,\ln (1 +
  x) + Li_{2}(-x) \right)  \,) \\ \nonumber
\end{eqnarray}
\noindent
For convenience the normalisation has been factored out, i.e. $N_c\,\alpha_{s}/\pi$ for a single quark flavor with unit charge and $N_c$ colours. Eq.(\ref{vacpol:1}) also determines $R$ for $x > 1$ through the symmetry relation: 

\begin{equation}
\label{vacpol:2}
R(x) = \left( 1 - {x^2} \right) \,R(0) + {x^2}\,R({1\over x})
\end{equation}
\noindent
The function $R(x)$ has the small-$x$ expansion:
\begin{equation}
\label{small-x}
R(x)\,/\,({{N_c^2-1} \over {N_c}})\,{3\over 8} = \hat{R}(x) = 
 1 + \left( \sum_{k = 2}^{\infty}a_{k}\,{x^k} \right)  + \left( \sum_{k
  = 3}^{\infty}b_{k}\,{x^k}\,\ln (x) \right) 
\end{equation}
\noindent
As a consequence, the following expansion for $D(Q^2)$ in terms of 
$M^2/Q^2$ can be derived:

\begin{eqnarray}
\label{D-expansion}
\lefteqn{D(Q^2) \,/\, ({{N_c^2-1} \over {N_c}})\,{3\over 8} = }  \\ \nonumber
& &1 + {{{M^2}}\over {{Q^2}}}\,2\,\left( \int _{0}^{1}dx\,{{\hat{R}(x) - 1}\over
  {{x^2}}} \right)  \\ \nonumber
& &+{\left( \,{{M^2}\over {Q^2}}\,\right)}^2\,\left( a_{2}\,(2\,\ln ({{{M^2}}\over
  {{Q^2}}}) + 1)  - 2 \int _{0}^{1}dx\,{{\hat{R}(x) - 1}\over x}
    - 2 \int _{0}^{1}dx\,{{\hat{R}(x)-1-a_{2}\,{x^2}}\over 
{{x^3}}}   \right)\nonumber \\  
& &+{{{\rm O}({{{M^2}}\over {{Q^2}}})}^3}  \\ \nonumber
\end{eqnarray}
\noindent
Numerically we have:

\begin{eqnarray}
\label{Koeff}
\lefteqn{  } \\ \nonumber
& &a_{2} = -1 \\ \nonumber
& &\int _{0}^{1}dx\,{{\hat{R}(x) - 1}\over x} = -7 + {{4\,{{\pi }^2}}\over 9} +
  2\,{\rm \zeta}(3) = -0.209 \\ \nonumber
& &\int _{0}^{1}dx\,{{\hat{R}(x) - 1}\over {{x^2}}} = {{-16}\over 3} + 4\,{\rm
  \zeta}(3) = -0.525 \\ \nonumber
& &\int _{0}^{1}dx\,{{\hat{R}(x) - 1 + {x^2}}\over {{x^3}}} = {{57 - 8\,{{\pi }^2}
  + 36\,{\rm \zeta}(3)}\over {18}} = 1.184 \\ \nonumber
\end{eqnarray}
\noindent 
Equation (\ref{D-expansion}) should be compared with the
result from the operator-product expansion:

\begin{equation}
\label{SVZ}
D(Q^2)={{N_c^2-1} \over {N_c}}\,{3\over 8} + {{\pi }\over {\alpha_{s}}}\,{{{2\,\pi }^2}\over N_c}\,
<{{\alpha_{s}\over {\pi}}\,GG}> {1\over {Q^4}} + O({{1}\over{Q^6}})
\end{equation}
The first term represents the perturbative contribution from
the diagrams in Fig.4 with massless gluons, and the second term is the
contribution from the gluon condensate as given in ref. \cite{SVZ}
(in our normalization).
Equating the $1/Q^4$ terms in (\ref{D-expansion}) and (\ref{SVZ}), we obtain:

\begin{equation}
\label{mass-cond:1}
<{{\alpha_{s}\over {\pi}}\,GG}>={{\alpha _{s}}\over {\pi}}\,
{{3(N_c^2-1)}\over {16\,{{\pi }^2}}}\,{M^4}\,\left( {{20}\over 3} -
  8\,\zeta (3) - 2\,\ln ({{{M^2}}\over {{Q^2}}}) \right)
\end{equation}
\noindent
The corresponding expression derived in \cite{Graz} reads:

\begin{equation}
\label{mass-cond:2}
<{{\alpha_{s}\over {\pi}}\,GG}>= {{\alpha_{s}\over {\pi}}\,
{{3(N_c^2-1)}\over {16\,\pi }^2}}\,{M^4}\,\left( {{45}\over 8} +
  {{10}\over 3}\,\zeta (3) - {{5\,{{\pi }^2}}\over {27}} - {5\over {18}}\,\ln
  ({{{M^2}}\over {{Q^2}}}) - {5\over 9}\,{{\ln ({{{M^2}}\over {{Q^2}}})}^2}
  \right)
\end{equation}
\noindent
which has no resemblance to what we have obtained.

At present the origin of this discrepancy is not satisfactorily explained.
However, since our own calculation only uses standard Feynman diagram 
techniques and has been checked by an independent calculation where the diagrams
in Fig.4 are calculated by a numerical procedure along the lines of 
ref. \cite{Kin}, we suspect that the Mellin transform technique used in ref. \cite{Graz}
has not been properly applied.

A $Q^2$-independent relation between M and the gluon condensate can be 
obtained from eq.(\ref{mass-cond:1}) by using for $\alpha_{s}/\pi$ the 
standard expression $4/\beta_0 ln(Q^2/\Lambda^2)$ for the running 
coupling with $\beta_0=N_c 11/3$ and evaluating  eq.(\ref{mass-cond:1}) 
to leading order in $ln(Q^2)$.
The result is:

\begin{equation}
\label{M_VP}
M^4 = {{22} \over {9}}\,\pi^2\,{{ N_c}\over {(N_c^2-1)}}\, 
<{{\alpha_{s}}\over {\pi}}\,GG>
\end{equation}
\noindent
yielding $M \simeq 570MeV$  for $N_c=3$ and the standard 
SVZ-value for the gluon condensate 
$<{{\alpha_{s}\over {\pi}}\,GG}>=(.33GeV)^4$.

\section{Scalar glue-ball current}

Instead of the electromagnetic current we consider in this section the
scalar glue-ball current defined in terms of the gluon field-strength
tensor $G^a_{\mu \nu}$ by

\begin{equation}
\label{GG-curr}
\hat{j} = G^a_{\mu \nu} G^{a \mu \nu}
\end{equation}
For ease of writing we use $\hat{j}$ to denote the current with a unit
normalization factor and  $j$ for the current with a normalization factor
$\alpha_s/\pi$ which turns $j$ into a renormalization 
group invariant quantity
\footnote{ For a leading logarithm analysis it will be sufficient to
use $\alpha_s/\pi$ instead of the more generally valid factor
$4\beta(g)/(\beta_0 g)$}.

As a first step we calculate the diagram in Fig. 5 for $\hat{j}$ 
in the massive theory. Defining $\hat{\pi}_{GG}(Q^2)$ for euclidean $q^2=-Q^2$
by:
 
\begin{equation}
\label{GG-pihat:1}
< \hat{j} \, \hat{j} > = i \, {{N_c^2-1} \over {2 \pi^2}}\,\hat{\pi}_{GG}(Q^2)
\end{equation}
we obtain: 
\begin{equation}
\label{GG-pihat:2}
\hat{\pi}_{GG}(Q^2) \cong \left( {1\over 2}\,{{\left( {Q^2} + 2\,{M^2} \right) }^2} + {M^4} \right)
  \,\left( \int _{0}^{1}d\alpha \,\ln (1 + {{\alpha \,\left( 1 - \alpha 
  \right) \,{Q^2}}\over {{M^2}}}) \right) 
\end{equation}
where $\cong$ means equality up to polynomial terms which do not survive
a three-fold subtraction needed to remove the UV-divergencies. From 
(\ref{GG-pihat:2}) we obtain the following expansion in terms of $M^2/Q^2$:

\begin{equation}
\label{GG-pihat:3}
Q^2 \, ({{d} \over {dQ^2}})^3 \,\hat{\pi}_{GG}  = 
1 - 3\,{{M^2}\over {Q^2}} + 12\,({{M^2}\over {Q^2}})^2 + {\rm
  O}({{M^2}\over {Q^2}})^3
\end{equation}
which is free of logarithmic factors $ln(M^2/Q^2)$ up to order $1/Q^4$.
As visible from this expression, the normalization of $\hat{\pi}_{GG}$ 
has been chosen such that the third derivative equals to one in the massless
limit.

\begin{figure}[htp]
\label{FGGcorr}
\epsfxsize=12cm
\epsffile{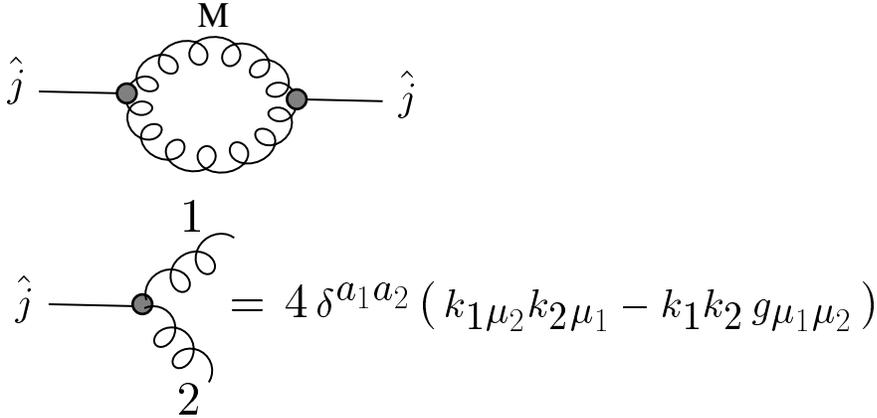}
\caption{The leading order diagram for the scalar glueball
current with a gluon of mass $M$, and the expression for the vertex.}
\end{figure}

As a second step we make a renormalization-group analysis for the 
coefficient function of the gluon condensate $<j>$ in the operator- 
product expansion of the
two-point function $\pi_{GG}$ for the scalar 
current $j=(\alpha_{s}/\pi)\,\hat{j}$:

\begin{equation}
\label{GG-pi:1}
\pi_{GG}  = {\rm C}_{pert}(Q^2) \,1 + {\rm C}_{GG}(Q^2)\,<j> + \, ...
\end{equation}
The coefficient function $C_{GG}$ has the following expansion in 
$\alpha_s/\pi$ :

\begin{equation}
\label{C_GG:1}
C_{GG}(Q^2) = {{\alpha _{s}({{\mu }^2})}\over {\pi }}\,
{\rm C}_0 + ({{{\alpha _{s}({{\mu}^2})}\over {\pi }}})^2\,
{\rm C}_1({{{{\mu }^2}}\over {{Q^2}}}) + \, ...
\end{equation}
where $C_0$ is a constant whose value can be found from applying Wick's
theorem to the product of currents $j(x)j(0)$:
\begin{equation}
\label{C_0}
C_0 = -\,{{8 \pi^2}\over{N_c^2-1}}
\end{equation}
Independence of $C_{GG}$ from the renormalization point $\mu$ requires the
coefficient $C_i$ in (\ref{C_GG:1}) to be a polynomial of degree $i$ in the
variable $ln(\mu^2/Q^2)$. Choosing $\mu^2 = Q^2$ and using the $1$-loop expression for $\alpha_s(Q^2)/\pi$, we obtain:

\begin{equation}
\label{C_GG:2}
Q^2 \, ({{d} \over {dQ^2}})^3 \,C_{GG}(Q^2)  = 
{{1} \over {(Q^2)^2}} \,( - 8\,{{C_0}\over {\beta_0 \, ln(Q^2/\Lambda^2)^2}} 
+ O(ln(Q^2/\Lambda^2)^{-3}) \, \, )
\end{equation}
where $\beta_0\,=\,11/3\,N_c$

The desired relation between M and the gluon condensate is now obtained
by matching the $1/Q^4$ terms in eq.(\ref{C_GG:2}) and 
$(\alpha_{s}(Q^2)/\pi)^2 \,\times \, eq.(\ref{GG-pihat:3})$.
The result is:

\begin{equation}
\label{M_GG}
M^4 = {{11}\over{9}}\,\pi^2\,{{N_c}\over{N_c^2-1}}\,
<{{\alpha_{s}\over {\pi}}\,GG}>
\end{equation}
Comparing with the result obtained in section 3 in eq.(\ref{M_VP}),
we recognize that we have almost obtained the same result except that
the numerical coefficient is smaller by a factor of $2$. Thus the
previous estimate for $M$ is reduced from $570MeV$ to $M \simeq 480 MeV$

It is interesting to compare eq.(\ref{M_GG}) with
the result of ref.(\cite{Kog}) obtained by a variational investigation
of an ansatz for the vacuum functional. Again both results have 
the same structure and color dependence (large $N_c$).
However, here there is a large difference in the numerical constant, i.e.
$120$ for ref.(\cite{Kog}) instead of $11/9$. As a consequence their mass estimate is larger by a factor $\simeq 3$.
\newpage
\section{ Summary and Outlook}
In this paper we have continued previous attempts to estimate 
values for the gluon mass by confronting results obtained in leading order calculations for massive gluons 
with experimental results; either by a direct comparison 
as for $J/\psi$ and $\Upsilon$ decays or in a more indirect way
by matching against nonperturbative contributions in the correlator
of the electromagnetic current or the scalar glue-ball current.
For the  decays of heavy quarkonia the Dalitz distributions 
involving $2-$ or $3-$ massive  gluons 
were derived, extending thus the basic formulae originally 
derived for positronium to this more general case. We then
analyzed the leading QCD correction of the electromagnetic
vacuum polarisation and determined the coefficients in an
expansion in $1/Q^2$ which are relevant for a comparison with the
leading nonperturbative contribution for this quantity. We
found a different result for the $1/Q^4$-coefficient as previously
given in the literature and claim that our result is correct as
it has been checked by a completely different method than the one
presented in this paper. Our result was then used to derive a
simple renormalization-group invariant expression between the 
gluon mass and the gluon condensate.  Finally a similar kind of analysis was
applied to the correlator of the scalar glue-ball current where 
gluon mass effects already appear on the $1$-loop level. Again this
leads to a relation between the gluon mass and the gluon condensate
which is renormalization group invariant although with a slightly
different numerical coefficient.

Taken at face value, there is a large spread 
between the mass estimates from
heavy quarkonia decays, in particular the $\Upsilon$-decays where
we can expect gluon mass effects to have fully developed,
and the values obtained from the analysis of the correlators.
Certainly the value
of $M \simeq 1.4GeV$ obtained from $\Upsilon$-decays should be 
considered as an 
upper bound, since the starting values of \cite{Cons} for the gluon mass
have been obtained under the assumption that the main suppression comes
from the gluon mass, not the relativistic corrections. As argued
for instance in \cite{Pir}, the pattern of suppressions in 
the total $J/\psi$ and $\Upsilon$ 
decay rates can quantitatively be reproduced if the momentum dependence
of the quarks is taken into account. Whether this straightforward
incorporation of relativistic effects is legitimate is, however, still
under debate (see \cite{Cons}), and therefore a substantial gluon mass -
though smaller in magnitude than the values discussed here - may 
indeed be required by the data. The results from matching the leading
nonperturbative correction in the correlator for the electromagnetic
and the scalar glueball current are equal to each other up to a factor 
of $2$ in the relation for $M^4$. Although this has little effect on the
numerical value of the gluon mass $M$, the dependence on the process
remains to be explained.
In this context it would also be
interesting to derive the expansion in powers of $M/M_{Q\bar{Q}}$
of the total decay rates of heavy quarkonia discussed in section 2
and match them likewise against the leading nonperturbative correction
from the gluon condensate.

A more complete analysis should also cover the $1/Q^2$ corrections
obtained in the massive theory, which seem to have no correspondence
in the operator-product expansion. It has been argued in ref.(\cite{Graz})
that these corrections cancel against corrections from the 
formation of a vortex condensate. We believe that this proposition
requires further examination.

\section{ Acknowledgement}
One of us (W.W.) would like to thank the Physics Department
of Wuhan University for the kind hospitality which has been
extended to him in 1995 and 1996 on the occasion of his visits
as foreign scholar.

\section{Appendix}

For the distribution of the photon in $^3S_1 \rightarrow \gamma+gg$ 
the following expression is obtained from
eq.(\ref{gamma2g}) by integrating over the gluons. 

\begin{eqnarray}
\label{gammadist}
\lefteqn{{{d\,\Gamma }\over {dz}} \propto } \\ \nonumber
& &{{x_{+}-x_{-}}\over {{z^2}}} \\ \nonumber
& & +\,{{ln(x_{+}/x_{-})}\over {{z^2}\,{{\left( -2 + 4\,\eta  + z \right) }^3}}}\,\lbrack
8\,{{\left( -1 + 2\,\eta  \right) }^2}\,\left( 2 - 4\,\eta  + 7\,{{\eta }^2}
  \right)  \\ \nonumber
& &\hspace{0pt}+8\,\left( -1 + 2\,\eta  \right) \,\left( 5 - 12\,\eta  +
  10\,{{\eta }^2} + 2\,{{\eta }^3} \right) \,z \\ \nonumber
& &\hspace{0pt}+2\,\left( -1 + 2\,\eta  \right) \,\left( -17 + 10\,\eta  +
  6\,{{\eta }^2} \right) \,{z^2} \\ \nonumber
& &\hspace{0pt}+2\,\left( -5 + 2\,\eta  + 2\,{{\eta }^2} \right) \,{z^3} \,\rbrack
  \\ \nonumber
& & +\,{{(1/x_{-}-1/x_{+})}\over {{z^2}\,{{\left( -2 + 4\,\eta  + z \right) }^2}}}\,\lbrack
4\,{{\left( -1 + 2\,\eta  \right) }^2}\,\left( 1 + 3\,{{\eta }^2} \right)  \\
  \nonumber
& &\hspace{0pt}+4\,\left( -1 + 2\,\eta  \right) \,\left( 3 - 4\,\eta  +
  2\,{{\eta }^2} + 2\,{{\eta }^3} \right) \,z \\ \nonumber
& &\hspace{0pt}+2\,\left( 7 - 18\,\eta  + 10\,{{\eta }^2} + 10\,{{\eta }^3}
  \right) \,{z^2} \\ \nonumber
& &\hspace{0pt}+4\,\left( 2 + \eta  \right) \,\left( -1 + 2\,\eta  \right)
  \,{z^3} + \left( 2 + \eta  \right) \,{z^4} \,\rbrack \\ \nonumber
\end{eqnarray}
Here $z=E_{\gamma}/m$ in terms of the photon energy $E_{\gamma}$, and 

\begin{equation}
x_{\pm} = 1 - 2\,\eta  - {z\over 2}  \pm {z\over 2}\,{\sqrt{1 - {{4\,\eta }\over {1
  - z}}}}
\end{equation}
delineate the boundary of the Dalitz plot in the plane $(x_{2},z)$ with 
$x_2= E_2 /m - 2\eta$.


\begin{thebibliography}{99}

\bibitem{Cor1} J.M. Cornwall, Nucl. Phys {\bf B157} (1979) 392

\bibitem{Cor2} J.M. Cornwall, Phys.Rev {\bf D26} (1982) 1453

\bibitem{SVZ} M.A. Shifman, A.I. Vainshtein, V.I Zakharov, Nucl.Phys. {\bf B147} (1979) 385

\bibitem{LLS} C.H. Llewellyn Smith, Phys.Lett {\bf 46B} (1973) 233,

J.M. Cornwall et al, Phys.Rev.Lett {\bf 30} (1973) 1268


\bibitem{Velt} M. Veltman, Acta Physica Polonia, {\bf B12} (1981) 437


\bibitem{Appl} T. Appelquist, C. Bernard, Phys.Rev {\bf D22} (1980) 200


\bibitem{Ak} R. Akhoury, Nucl.Phys. {\bf B234} (1984) 533

\bibitem{tH1} G. t'Hooft, Nucl.Phys {\bf B138} (1978) 1, ibid B153 (1979) 141

\bibitem{Par} G. Parisi, R. Petronzio, Phys. Lett {\bf 94B} (1980) 51

\bibitem{Soni} J.M. Cornwall, A. Soni, Phys. Lett {\bf 120B} (1983) 431 

\bibitem{Szcz} A. Szczepaniak, E.S. Swanson, C.-R. Ji, S.R. Cotanch, 
 Phys. Rev. Lett.{\bf 76} (1996) 2011 (hep-ph/9511422)

\bibitem{Graz} F.R. Graziani, Z.Phys. {\bf C33} (1987) 397

\bibitem{Halz} F. Halzen, G. Krein, A.A. Natale, Phys.Rev {\bf D47} (1993) 295

\bibitem{Kog} I.I. Kogan, A. Kovner, Phys. Rev. {\bf D52} (1995) 3719, (hep-th/9408081)

 
\bibitem{Cons} M. Consoli, J.H. Field, Univ. Geneva preprint UGVA-DPNC 
1994/12-164

\bibitem{LandLif} Landau-Lifschitz, Quantum Electrodynamics

\bibitem{Wet} W. Wetzel, W. Bernreuther, Phys.Rev. {\bf D24} (1981) 2724

\bibitem{Kin} P. Cvitanovic, T. Kinoshita, Phys.Rev. {\bf D10} (1974) 3978, 3991

\bibitem{Pir} H.C. Chiang, , J. H\"ufner, H.J. Pirner, 
 Phys. Lett {\bf 324B} (1994) 482

\end{thebibliography}
\end{document}